\documentclass[12pt]{iopart}
\usepackage{epsf}
\begin{document}

\title{Rock-scissors-paper game on regular small-world networks}
\author{Gy\"orgy Szab\'o$\dag$, Attila Szolnoki$\dag$ and
Rudolf Izs\'ak$\ddag$}
\address
{$\dag$Research Institute for Technical Physics and Materials Science,
P.O. Box 49, H-1525 Budapest, Hungary}
\address
{$\ddag$Institute for Theoretical Physics, E\"otv\"os University, 
P.O. Box 32, H-1518 Budapest, Hungary}

\date{\today}

\begin{abstract}
The spatial rock-scissors-paper game (or cyclic Lotka-Volterra system)
is extended to study how the spatiotemporal patterns are affected by the
rewired host lattice providing uniform number of neighbours (degree)
at each site. On the square lattice this system exhibits a self-organizing
pattern with equal concentration of the competing strategies (species).
If the quenched background is constructed by substituting random links
for the nearest neighbour bonds of a square lattice then a limit cycle
occurs when the portion of random links exceeds a threshold value. This
transition can also be observed if the standard link is replaced temporarily
by a random one with a probability $P$ at each step of iteration. Above a
second threshold value of $P$ the amplitude of global oscillation increases
with time and finally the system reaches one of the homogeneous (absorbing)
states. In this case the results of Monte Carlo simulations are compared
with the predictions of the dynamical cluster technique evaluating all
the configuration probabilities on one-, two-, four- and six-site
clusters.
\end{abstract}

%\pacs{05.50.+q, 87.23.Cc}


\maketitle

\section{Introduction}

The rock-scissors-paper like cyclic dominance among three states (modes, 
species, strategies, opinions) are widely studied in different spatial 
systems. For example, the Rayleigh-Bernard convection in fluid layers
rotating around a vertical axis exhibits the K\"upper-Lortz instability
\cite{kupper:jfm69} resulting in a cyclic change of the three possible
directions of parallel convection rolls
\cite{busse:sci80,toral:pa00,gallego:pre01}. 
Such a situation can appear in biological (ecological) systems too
\cite{may:siam75,buss:an79,tainaka:epl91,sinervo:nature96}.
Very recently, Kerr {\it et al.} \cite{kerr:nature02} have
justified experimentally that the cyclic dominance between the toxic,
sensitive and resistant microbes maintains their coexistence as
predicted previously by several theoretical works
\cite{nakamaru:tpb00,szabo:pre01a,czaran:pnas02}. The emergence of cyclic
invasions has also been observed for some three-strategy evolutionary games
where this phenomenon promotes the cooperation among selfish individuals
\cite{szabo:pre00a,hauert:science02,szabo:prl02}.

The above three-state systems exhibit some general features. Namely, spiral 
formation (or rotating three-edge vortices and antivortices) can occur
on the two-dimensional backgrounds \cite{sole:pla92,tainaka:pre94,
szabo:pre02a}
as it is observed for the Belousov-Zhabotinskii reaction 
as well as for the numerical investigation of excitable 
activator-inhibitor media \cite{hempel:prl99}. This self-organizing
structure can provide a stability against some external
invaders and thereby it plays crucial role in ecological systems
\cite{boerlijst:pd91,szabo:pre01b}. Furthermore the cyclic dominance yields
a paradoxical response to the external support of one of the species
\cite{tainaka:pla93,frean:prs01} and global oscillation can occur
when varying either the model \cite{szabo:pre02d} or the structural
parameters \cite{kuperman:prl01} if long range interactions are allowed.

In this work the spatial rock-scissors-paper game will be extended for
such regular networks where each site has four neighbours. We discuss how
the quenched and annealed randomness of the network affects the 
spatiotemporal distribution of species. Such a comparison has already been
performed for a rumor propagation model \cite{zanette:pre01,zanette:pre02}.
Now, the randomness is introduced in small-world manner \cite{watts:nature98}
leaving the degree of sites unchanged. This restriction does not
influence the small-world feature of the network and it simplifies the
application of some theoretical approximation at least for the case of
annealed randomness. However, the main conclusion remains valid for both
types of randomness. The Monte Carlo (MC) simulations indicate that 
these systems undergo a transition (Hopf bifurcation) from
a stationary state (with fixed average concentrations) to a global
oscillation. The amplitude of oscillation increases with the measure of
randomness and the increasing oscillation terminates at one of the
absorbing (homogeneous) states above a second threshold value for
annealed randomness. We show that neither the mean-field nor the pair 
approximations can explain these transitions. The failure of these
descriptions has inspired us to use the more sophisticated dynamical
cluster techniques where one determines all the configuration probabilities
on a $k$-site cluster. At the levels of $k=1$ and 2 this technique is
equivalent to the mentioned mean-field and pair approximations. The essence
of this simple method is described in Refs. \cite{marro:99,dickman:pre01}.

\section{The model}

We consider a very simple model where the sites of a regular graph
are occupied by one of the three species ($s_i=1, 2, 3$) that dominate
cyclically each other. The evolutionary process is governed by the sequence
of elementary invasions along the randomly chosen edges of the graph. 
Namely, first we choose a site and one of its linked (neighbouring) sites 
at random. If these two sites are
occupied by different species then the predator occupies the prey's site,
i.e., the $(1,2)$ and $(2,1)$ pairs transform into $(1,1)$, $(2,3)$ and 
$(3,2)$ into $(2,2)$, and $(1,3)$ and $(3,1)$ into $(3,3)$.
Starting from a random initial state the above process is repeated until
the system reaches the stationary state or limit cycle we study.
This system was already analysed systematically by several authors if the
sites form a $d$-dimensional lattice \cite{tainaka:pre94,frachebourg:jpa98}.

\begin{figure}[h]
\begin{center}
             \epsfxsize=7cm
             \epsfbox{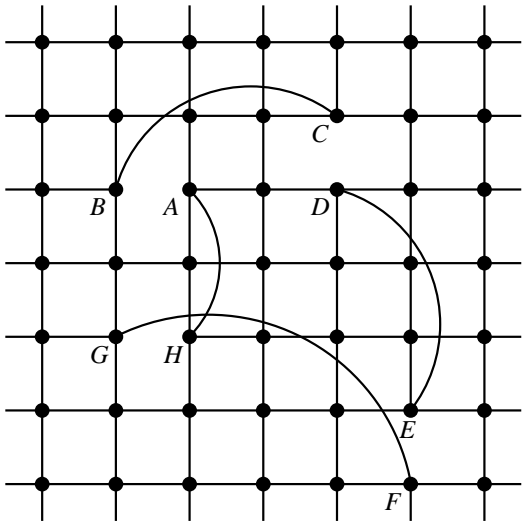}
\end{center}
\caption{Structure of a regular small-world network whose
construction starts from a square lattice. First the randomly chosen $AB$
link is removed and the site $B$ is rewired to the randomly chosen site $C$.
To have four connections at site $C$ we eliminate one of the previous
links (here $CD$) and we add a new link $DE$ at random. This process
is repeated until $Q$ portion of the nearest neighbour bonds are replaced
by random links. Finally the last site (here $H$) is wired to the first
one ($A$). The edges pointing out along the periphery refer to periodic
boundary conditions assumed for the original square lattice.}
\label{fig:rswstruc}
\end{figure}

Two types of random structures will be contrasted with each other. In the
first case the structure is quenched after the generation of edges.
Figure \ref{fig:rswstruc} illustrates an example whose creation is similar
to those suggested by Watts and Strogatz \cite{watts:nature98}. Notice,
that the present algorithm (explained in the caption of
Fig.~\ref{fig:rswstruc}) conserves the degree of sites, i.e., the number
of neighbours remains fixed ($z=4$) for each site. In the second case the
``neighbourhood'' is not fixed in time, that is a randomly
chosen new site can be replaced for any standard neighbours during the
elementary invasions explained above. These types of random networks
can characterize the interaction among individuals in social systems
\cite{szabo:pre02d,watts:nature98,zimmermann:01,abramson:pre01,ebel:pre02}.
In both cases the random links connect two sites being arbitrary distance
from each other in the original structure.

The measure of quenched randomness is characterized by the $Q$ portion of
random links substituted for the nearest neighbour bonds. If $Q=0$ this
structure reproduces the square lattice and for $Q=1$ it is equivalent to a
random regular graph \cite{bollobas:95} where the typical local structure
is analogous to a Cayley tree for large number of sites $N$.
Previous investigations indicated that some phenomena on the random regular 
graphs can be well described analytically if the background is assumed
to be a Bethe lattice in the limit $N \to \infty$
\cite{szabo:pre02d,szabo:pre00b}.  In other words, the structure with
$Q=1$ can be considered as a possible realization of the Bethe lattice
in the simulations for large $N$.

Unfortunately, the present random regular structures with $0<Q<1$ are not 
yet  investigated rigorously, although many other classes of networks are
well studied \cite{amaral:pnas00,albert:rmp02,dorogovtsev:03,newman:siam03}.
We think that the constraint of regularity leaves the relevant features
unchanged and the present structure remains similar to those introduced by
Watts and Strogatz \cite{watts:nature98} on a square lattice. When
increasing $Q$ the present structure exhibits a continuous transition
from the square lattice to the random regular graph. 

Besides the above quenched randomness we will investigate the consequences
of the annealed (temporal) randomness in the structure. In this case the
standard links are defined by the bonds between the nearest neighbours
sites forming a square lattice. Occasionally the standard link is replaced by
a random one with a probability $P$ characterizing the strength of annealed
randomness. Evidently, for $P=0$ the structure is equivalent to 
the square lattice. On the other hand, in the limit $P \to 1$ this system 
satisfies the conditions of mean-field approaches.

\section{Simulations}

The MC simulations are performed on a network consisting of $N=L \times L$
sites where the linear size of the square lattice ($L$) is varied from
400 to 3200.
The regular small-world networks are constructed from a square
lattice repeating the rewiring steps $-2N \ln (1-Q)$ times as explained in
the caption of Fig.~\ref{fig:rswstruc}. (The logarithmic correction
takes into account that the substitution of a random link for another
random one does not increase the quenched randomness.) 

The above model has two parameters characteristic to the quenched
($Q$) and annealed ($P$) randomness of the background. The present analysis
is restricted to those cases when one of them is chosen to be zero.

For small sizes ($L<10$) this system evolves into
one of the three absorbing states where all the sites are occupied by
the same species. For sufficiently large system sizes, however, the three
species can coexist and after some transient time the state can be well
described by the current concentration of the three species
[$c_1(t)+c_2(t)+c_3(t)=1$].
In order to observe these states we have to choose such a large $L$ where
the amplitude of fluctuations becomes significantly less than the minimum
value of concentrations. The above choices satisfy this condition.

\begin{figure}[h]
\begin{center}
             \epsfxsize=8cm
             \epsfbox{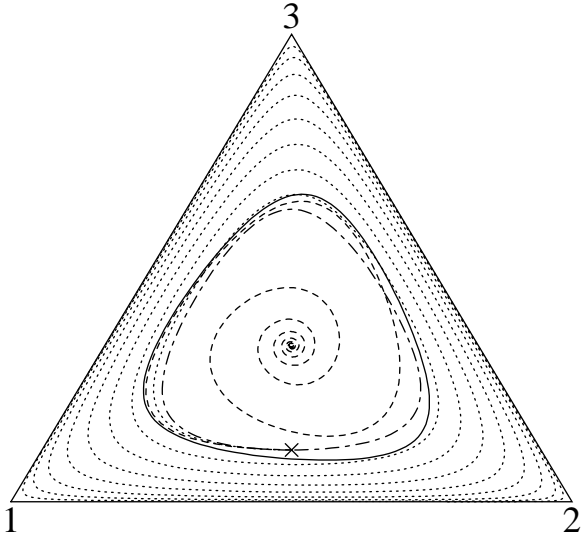}
\end{center}
\caption{The MC simulations show four typical trajectories
plotted on the ternary phase diagram. All the simulations are started from
the same initial state (denoted by symbol X) on the square lattice for
$L=3200$. If $P=0$ then the system develops into the central fixed point
(dashed line). For $P =0.2$ the growing spiral trajectory (dotted line) 
ends at one of the homogeneous state. The solid line indicates only the
limit cycle toward the growing (or shrinking) spiral trajectories evolve
for $P=0.05$. In the mean-field limit ($P=1$) the trajectory (dash-dotted
line) returns periodically to the initial state.}
\label{fig:evol}
\end{figure}

On the square lattice ($Q=P=0$) this system develops into a stationary
state where all three species are present with the same average
concentration, {\it i.e.},
$\langle c_1 \rangle = \langle c_2 \rangle = \langle c_3 \rangle =1/3$
corresponding to the central point in the ternary phase diagram as plotted
in Fig.~\ref{fig:evol}. In this case the three species alternate cyclically
at each site and the short range interactions are not able to synchronize
these local oscillations. 

The corresponding self-organizing spatiotemporal patterns are well
investigated previously by several authors
\cite{tainaka:pre94,frachebourg:jpa98,szabo:pre99}. In these patterns
the rotating vortices (spirals) and antivortices are not recognizable
because of the interfacial roughening. The absence of smooth interfaces
(surface tension) is caused by the fact that the elementary invasions
between two neighbouring sites are not affected by their neighbourhood
\cite{szabo:pre02a}.

\begin{figure}[h]
\begin{center}
             \epsfxsize=8cm
             \epsfbox{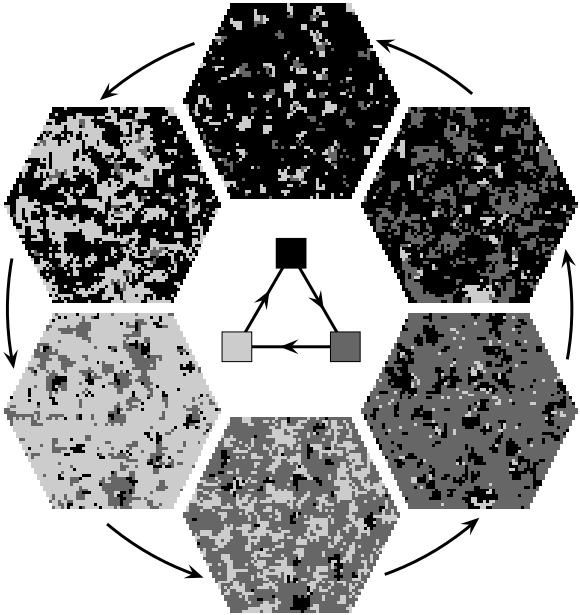}
\end{center}
\caption{The haxagonal snapshots represent consecutive species
distributions on a small portion of square lattice during the MC simulation
of global oscillation for $Q=0$ and $P=0.1$. Arrows along the periphery
indicate the direction of time evolution. The cyclic food web is shown in
the center where the three species are denoted by different gray scales
as in the snapshots and the arrows point to the direction of invasion
between the species. Thus, the territory of the "black" species will
be occupied by the "light-gray" species. Once the "ligh-gray"s prevail,
they will be invaded by the "dark-gray" species, and finally a "black"
invasion closes the cycle as shown by the snapshots.}
\label{fig:osc6ff}
\end{figure}

Global oscillation (synchronization) occurs when the frequency of random
(long range) links exceeds a threshold value dependent on whether
quenched or annealed randomness is considered. To illustrate the
time-dependence of the species distributions during a period of this
global oscillation six consecutive snapshots are plotted
in Fig.~\ref{fig:osc6ff}.

The amplitude of oscillation increases with the measure of randomness
in both cases.
If the annealed randomness exceeds a second threshold value then the 
trajectories approach the edges of the triangle and sooner or later the
evolution is terminated at one of the homogeneous (absorbing) states 
(where the system stays for ever).
It is emphasized, however, that the homogeneous states are not stable
against the invasion of their predators. For example, the offsprings
of a single species 3 will occupy the whole territory of species 1
in the absence of species 2.

In the ternary phase diagram the shape of the limit cycles reflects the
cyclic symmetry between the species and its extension is described by the
area $A$ compared to its maximum value. The average value of this 
quantity can be easily determined by numerical integration after a
suitable relaxation time for either the MC simulations or the
dynamical cluster techniques.
Evidently, $A$ vanishes if the system tends toward the central fixed 
point, and it goes to 1 when the trajectories approach the edges
of triangle (see Fig.~\ref{fig:evol}).

\begin{figure}[h]
\begin{center}
             \epsfxsize=8cm
             \epsfbox{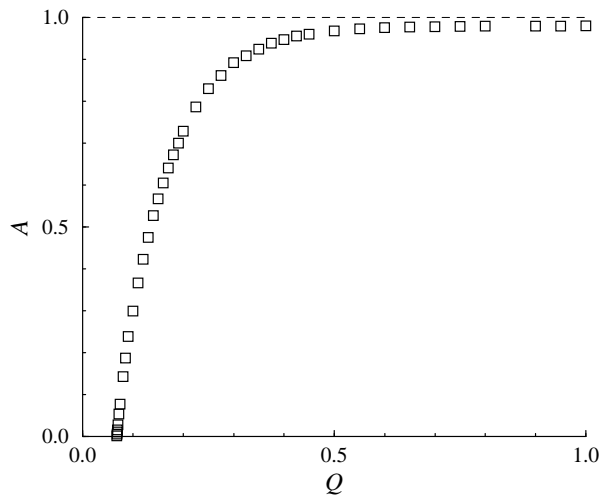}
\end{center}
\caption{Relative area of limit cycle versus $Q$
for a regular small-world system sketched in Fig. \ref{fig:rswstruc}.}
\label{fig:aq}
\end{figure}

Systematic MC simulations are carried out to determine the average value
of $A$ on the regular small-world structure for different $Q$ values.
The results in Fig.~\ref{fig:aq} refer to a transition from a fixed point
to the limit cycle. For weak randomness [$Q < Q_1=0.067(1)$] the system
always tends to the central fixed point. Conversely, oscillating behaviour
occurs and the area $A$ (as well as the amplitude) of the limit cycle
increases monotonously with $Q$ and tends to the value $A(Q=1)=0.980(1)$.
The limit $Q \to 1$ is investigated on a random regular graph created by
using a different algorithm \cite{szabo:pre00b}. In the vicinity of the
transition point $A$ vanishes linearly with $Q - Q_1$ in agreement with
the prediction of Hopf bifurcation describing the emergence of a limit
cycle in a mean-field type system if the model parameters are varied. 
Further rigorous investigations are required to quantify the variations
in the spatiotemporal patterns when approaching the transition point.

As it is already mentioned the system size should be large enough to
avoid the finite size effects. The sharpe transition to the global 
oscillation is wiped out by fluctuations on smaller systems as
discussed previously by Kuperman and Abramson \cite{kuperman:prl01} who
considered a three-state (cyclic) epidemiological model on small world
networks suggested by Watts and Strogatz \cite{watts:nature98}.
The finite size effects are more dangerous when the value
of area $A$ approaches 1. In this case the evolution may be easily traped
by one of the absorbing states. This difficulty is avoided by choosing
sufficiently large system size. As a result, the plotted $A(Q)$ function
may be considered as the infinite size limit.

\begin{figure}[h]
\begin{center}
             \epsfxsize=8cm
             \epsfbox{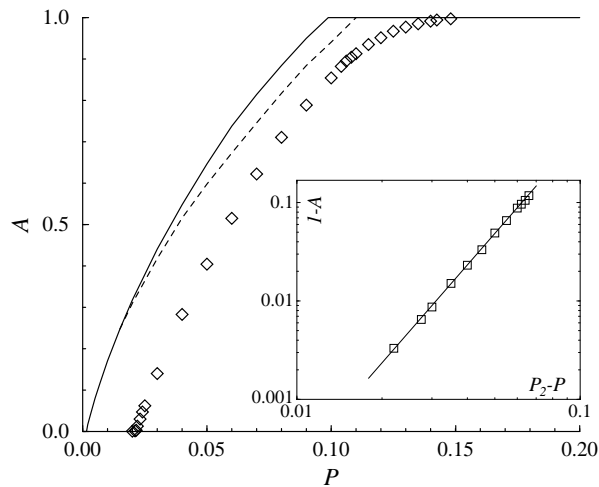}
\end{center}
\caption{Relative area of limit cycle as a function of $P$
characteristic to the annealed randomness. The symbols indicate the MC
data, the solid and dashed lines illustrate the prediction of four-
and six-site dynamical cluster techniques. The inset shows the log-log plot
of $1-A$ {\it vs.} $P_2-P$.}
\label{fig:ap}
\end{figure}

The above analysis was repeated for the annealed randomness when varying
$P$ for $Q=0$. The results of MC simulations are summarized in 
Fig.~\ref{fig:ap}. The oscillating behaviour can be observed
for $P_1 < P < P_2$ where $P_1=0.020(1)$ and $P_2=0.170(1)$. If $P > P_2$
then the increasing spiral trajectory terminates in one of the absorbing 
states as demonstrated in Fig.~\ref{fig:evol}. In the vicinity of the first
threshold value $A \propto (P-P_1)$ in agreement with the expectation.
On the contrary, $A$ approaches 1 very smoothly when $P$ goes to $P_2$.
Surprisingly, our MC data can be well approximated by a power law behaviour
[$1-A \propto (P_2-P)^{\gamma}$] as indicated in the inset of
Fig.~\ref{fig:ap}. The numerical fit yields $\gamma = 3.3(3)$.

It is worth mentioning that the emergence of global oscillation has
already been observed in the above mentioned epidemiological model when 
varying the quenched randomness without the consraints of regularity
\cite{kuperman:prl01}. The similar behavior refers to the robustness
of this type of transition. 

\section{Extended mean-field analysis}

The predictions of the traditional mean-field analysis are well discussed
in the textbook by Hofbauer and Sigmund \cite{hofbauer:98}. According to
this approach four
stationary states exist, namely, the above mentioned three absorbing states
(e.g. $c_1=1$ and $c_2=c_3=0$) and the well-mixed symmetric state where
the three species are present with the same concentration (1/3). Besides
these stationary solutions the mean-field analysis shows the existence
of set of oscillating states whose closed trajectories draw concentric
orbits around the centrum in the ternary phase diagram. Along these
orbits the product $c_1 c_2 c_3$ is conserved. In agreement
with the expectation the MC simulations reproduce this behaviour for
$P=1$ as shown in Fig.~\ref{fig:evol}.

The application of the pair approximation is inspired by its success
when investigating an evolutionary game with three (cyclically dominated)
strategies on a random regular graph \cite{szabo:pre02d}. This model 
exhibits both transitions mentioned above when varying the parameter(s)
of payoff matrix. It is underlined that here the adoption of the
neighbouring strategies depends on the neighbourhood.

In the pair approximations one determines the $p_2(s_1,s_2)$ probability
of finding $(s_1,s_2)$ configuration on two nearest neighbour (or linked) 
sites. In this case equations of motion are derived for these quantities
taking into account the contribution of all the elementary invasion
processes (details are given in the textbook by Marro and Dickman 
\cite{marro:99}). The numerical integration of the corresponding equations
predicts growing spirals approaching the boundaries (and resulting in
the maximum value $A=1$ in the limit $t \to \infty$) as indicated
in Fig.~\ref{fig:evol}. This prediction is contrary to the results of
MC simulations obtained on either the square lattice ($A=0$) or the
random regular graphs ($A=0.98$). We have to emphasize that although the
equations of motion involve explicitly the number of neighbours the simple
pair approximation can not distinguish the structure of the Bethe 
and square lattices. In the light of previous experiences \cite{szabo:pre02d}
it is expected that the pair approximation can well describe the behaviour
observed by MC simulations on the random regular graph because in this
structure the average loop size increases with with $\ln N$
\cite{bollobas:95}. Thus, for large $N$, the local structure is tree-like
and the short range correlations between two sites can be well approximated
by a product of the $p_2(s,s^{\prime})$ configuration probabilities involved
along the single path connecting the two sites. The comparison of the
above values of $A$ does not confirm this expectation. More precisely,
the pair approximation can not account for the effect preventing the
unlimited growth of the area of limit cycle. Preliminary results
suggest that the limit value of $A$ depends on the degree of the 
random regular graph. In the near future we wish to study this effect
by a suitable extension of the dynamical cluster techniques. Henceforth,
however, our efforts will be concentrated on the annealed randomness
because its investigation can be easily builted into the dynamical cluster
technique. At the level of pair approximation the corresponding results
predict that the "spiral pitch" decreases when $P$ is increased and
vanishes in the mean-field limit ($P=1$) as expected.

On the square lattice, as mentioned above, the pair approximation is not
capable to describe the appearance of self-organizing patterns maintained
by the interfacial motions due to cyclic invasions. This striking shortage
can be eliminated by choosing larger and larger clusters on which we
determine all the possible configuration probabilities \cite{marro:99}.
For example, at the level of four-site approximation we determine the
configuration probabilities $p_4(s_1,s_2,s_3,s_4)$ on a $2 \times 2$
cluster assumed to be translation invariant on the square lattice.
A recent summary of larger-cluster approximations can be found in
Ref.~\cite{dickman:pre01}.

\begin{figure}[h]
\begin{center}
             \epsfxsize=4cm
             \epsfbox{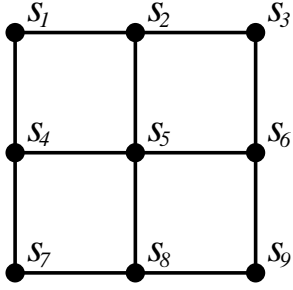}
\end{center}
\caption{The invasion of the central site from its neighboring sites affects
all the four four-site configuration probabilities.}
\label{fig:invcf}
\end{figure}

This approach takes explicitly into account some topological features of
the square lattice. A nearest-neighbor invasion
(e.g. $s_4 \to s_5$ as demonstrated in Fig.~\ref{fig:invcf}) will
simultaneously affect all the four four-site configuration probabilities
involved. The spatial effect is taken into account more rigorously if
the probability of such a nine-site constellation is approximated as
\begin{eqnarray}
p_9(s_1, \ldots ,s_9)=&&{p_4(s_1,s_2,s_4,s_5) p_4(s_2,s_3,s_5,s_6)
\over p_2(s_2,s_5) p_2(s_4,s_5)} \nonumber \\
&& \label{eq:p9} \\
&&\times {p_4(s_4,s_5,s_7,s_8) p_4(s_5,s_6,s_8,s_9)
\over p_2(s_5,s_6) p_2(s_5,s_8)} p_1(s_5)\;, \nonumber
\end{eqnarray}
where the configuration probabilities satisfy the following compatibility
conditions:
\begin{eqnarray}
p_1(s_1)=&&\sum_{s_2}p_2(s_1,s_2)=\sum_{s_2}p_2(s_2,s_1) \; ,  \nonumber \\
\label{eq:comp} \\
p_2(s_1,s_2)&=&\sum_{s_3,s_4}p_4(s_1,s_2,s_3,s_4)=
             \sum_{s_3,s_4}p_4(s_3,s_4,s_1,s_2)  \nonumber \\
             &=&\sum_{s_3,s_4}p_4(s_1,s_3,s_2,s_4)=
             \sum_{s_3,s_4}p_4(s_3,s_1,s_4,s_2) \; . \nonumber 
\end{eqnarray}

The time derivative of $p_4(s_1,s_2,s_3,s_4)$ can be described by a master
equation that summarizes the contribution of all the possible elementary 
invasions. Namely,
\begin{eqnarray}
{d \over dt} p_4(s_1,s_2,s_3,s_4)=&& \nonumber \\
-P \sum_{s_x} p_4(s_1,s_2,s_3,s_4) p_1(s_x)[\Gamma (s_x,s_1)+\Gamma (s_x,s_2)+
\Gamma (s_x,s_3)+\Gamma(s_x,s_4)] \nonumber \\
+P \sum_{s_x} p_4(s_x,s_2,s_3,s_4) p_1(s_1) \Gamma (s_1,s_x) \nonumber \\
+P \sum_{s_x} p_4(s_1,s_x,s_3,s_4) p_1(s_2) \Gamma (s_2,s_x) \nonumber \\
+P \sum_{s_x} p_4(s_1,s_2,s_x,s_4) p_1(s_3) \Gamma (s_3,s_x) \nonumber \\
+P \sum_{s_x} p_4(s_1,s_2,s_3,s_x) p_1(s_4) \Gamma (s_4,s_x) \label{eq:mast} \\
-{1-P \over 4} \sum_{s_5,s_6,s_7,s_8,s_9}p_9(s_1,s_2,s_5,s_3,s_4,s_6,s_7,s_8,s_9)
\Gamma (s_2,s_4) \nonumber \\
-{1-P \over 4} \sum_{s_5,s_6,s_7,s_8,s_9}p_9(s_5,s_1,s_2,s_6,s_3,s_4,s_7,s_8,s_9)
\Gamma (s_1,s_3) \nonumber \\
-{1-P \over 4} \sum_{s_5,s_6,s_7,s_8,s_9}p_9(s_5,s_6,s_7,s_1,s_2,s_8,s_3,s_4,s_9)
\Gamma (s_6,s_2) \nonumber \\
-{1-P \over 4} \sum_{s_5,s_6,s_7,s_8,s_9}p_9(s_5,s_6,s_7,s_8,s_1,s_2,s_9,s_3,s_4)
\Gamma (s_6,s_1) \nonumber \\
+{1-P \over 4} \sum_{s_x,s_5,s_6,s_7,s_8,s_9}p_9(s_1,s_2,s_5,s_3,s_x,s_6,s_7,s_8,s_9)
\delta (s_2,s_4) \Gamma (s_2,s_x) \nonumber \\
+{1-P \over 4} \sum_{s_x,s_5,s_6,s_7,s_8,s_9}p_9(s_5,s_1,s_2,s_6,s_x,s_4,s_7,s_8,s_9)
\delta (s_1,s_3) \Gamma (s_1,s_x) \nonumber \\
+{1-P \over 4} \sum_{s_x,s_5,s_6,s_7,s_8,s_9}p_9(s_5,s_6,s_7,s_1,s_x,s_8,s_3,s_4,s_9)
\delta (s_6,s_2) \Gamma (s_6,s_x) \nonumber \\
+{1-P \over 4} \sum_{s_x,s_5,s_6,s_7,s_8,s_9}p_9(s_5,s_6,s_7,s_8,s_x,s_2,s_9,s_3,s_4)
\delta (s_6,s_1) \Gamma (s_6,s_x) \nonumber \\
+ \ldots \nonumber 
\end{eqnarray}
where $\delta (s_x,s_y)$ denotes the Kronecker's delta and the constraint of invasion
is expressed as
\begin{equation}
\Gamma (s_x,s_y) = \cases{1, & if $s_y-1=s_x\; \mbox{mod}\; 3$ ; \cr
           0, & otherwise . \cr}
\label{eq:gamma}
\end{equation}
The terms proportional to $P$ describe the contributions coming from
the invasions from arbitrary distance while the contributions from one 
of the four nearest-neighbor sites are proportional to $(1-P)/4$. Equation 
(\ref{eq:mast}) involves explicitly only those terms coming from the downward
invasions. The derivation of the missing terms is straightforward.

At the level of six-site approximation the probability of a nine-site
configuration (as shown in Fig. \ref{fig:invcf}) is expressed
by the product of configuration probabilities on $3 \times 2$ clusters as
\begin{equation}
p_9(s_1, \ldots,s_9)={p_6(s_1,s_2,s_3,s_4,s_5,s_6)
 p_6(s_4,s_5,s_6,s_7,s_8,s_9) \over p_3(s_4,s_5,s_6)}
\label{eq:p96}
\end{equation}
where the $p_3(s_1,s_2,s_3)$ indicates the configuration probabilities
on a $3 \times 1$ cluster. Evidently, in this case the invasion of the
central site will influence some other six-site configuration probabilities
that we can handle in a similar way.

In both cases the corresponding master equations are solved by numerical
integration and the results are summarized
in Fig.~\ref{fig:ap}. In agreement with the expectation, the dynamical 
cluster techniques (at such a high level) reproduce qualitatively well
the results obtained by MC simulations. Namely, both descriptions
confirm the stability of the central stationary state, that is the area 
$A$ tends to zero, if $P<P_1^{(4p)}=P_1^{(6p)}=0.011(1)$. Above this
transition point the present approaches predict the appearance of a limit
cycle within a range $P_1<P<P_2$. The area $A$ increases linearly with
$P-P_1^{(4p)}$ in the close vicinity of the transition point. Similarly,
$A$ approaches to 1 linearly for both the four- and six-site approximations,
although these methods predict different transition points, i.e.,
$P_2^{(4p)}=0.097(3)$ and $P_2^{(6p)}=0.109(3)$. These sophisticated
techniques have significantly improved the description and the deviations
from the MC results reflect the relevant role of topological features.

\section{Conclusions}

In summary, cyclic invasions like the rock-scissors-paper in the
three-state systems can maintain a rich variety of spatiotemporal patterns
that depend on the quenched and annealed randomness of the background.
According to the MC simulations a self-organizing pattern occurs on the
square lattice. The classical mean-field and pair approximations are not
capable to reproduce this behaviour. It is demonstrated, however, that the
extended versions of this approach, called dynamical cluster technique at
the level of four- and six-site approximations, can describe the appearance
of this self-organizing pattern.

The quenched randomness is generated by a modified rewiring technique 
that conserves the degree at each site. For weak quenched
randomness the above spatiotemporal pattern remains stable.
When the measure of quenched randomness exceeds a threshold value
this system undergoes a transition from the symmetric stationary state
(central fixed point) to a synchronized oscillation (limit cycle).
For annealed randomness this model exhibits similar behaviour with a higher
sensitivity to the variation of annealed randomness and above a second
threshold value the increasing oscillation terminates at one of the
homogeneous (absorbing) states. These features are reproduced 
qualitatively well by the dynamical cluster technique considering the
configuration probabilities on four- and six-site clusters. 
We think that further systematic analyses can clarify how the transitions
are affected on those systems where the quenched and annealed randomness
occur simultaneously. More significant modifications of this technique
are required if we wish to study the cyclic invasions on networks with
arbitrary degree distributions.

\section{Acknowledgement}
This work was supported by the Hungarian National Research Fund under
Grant Nos. T-33098, F-30499 and Bolyai Grant No. BO/0067/00.

\end{document}